\documentclass{ws-ijmpcs}

\begin{document}

\markboth{L. Lavezzi et al. (on behalf of the CGEM-IT group)}
{The new cylindrical GEM inner tracker of BESIII}

%
\catchline{}{}{}{}{}
%

\title{The new Cylindrical GEM Inner Tracker of BESIII}

\author{L. Lavezzi$^{a,f,}$\footnote{corresponding author, {\it lia.lavezzi@to.infn.it, lavezzi@ihep.ac.cn}}, M. Alexeev$^{f}$, A. Amoroso$^{f,l}$, R. Baldini Ferroli$^{a,c}$, M. Bertani$^{c}$, D. Bettoni$^{b}$, F. Bianchi$^{f,l}$, A. Calcaterra$^{c}$, N. Canale$^{b}$, M. Capodiferro$^{c,e}$, V. Carassiti$^{b}$, S. Cerioni$^{c}$, JY. Chai$^{a,f,h}$, S. Chiozzi$^{b}$, G. Cibinetto$^{b}$, F. Cossio$^{f,h}$, A. Cotta Ramusino$^{b}$, F. De Mori$^{f,l}$, M. Destefanis$^{f,l}$, J. Dong$^{c}$, F. Evangelisti$^{b}$, R. Farinelli$^{b,i}$, L. Fava$^{f}$, G. Felici$^{c}$, E. Fioravanti$^{b}$, I. Garzia$^{b,i}$, M. Gatta$^{c}$, M. Greco$^{f,l}$, CY.Leng$^{a,f,h}$, H. Li$^{a,f}$, M. Maggiora$^{f,l}$, R. Malaguti$^{b}$, S. Marcello$^{f,l}$, M. Melchiorri$^{b}$, G. Mezzadri$^{b,i}$, M. Mignone$^{f}$, G. Morello$^{c}$, S. Pacetti$^{d,k}$, P. Patteri$^{c}$, J. Pellegrino$^{f,l}$, A. Pelosi$^{c,e}$, A. Rivetti$^{f}$, M. D. Rolo$^{f}$, M. Savri\'e$^{b,i}$, M. Scodeggio$^{b,i}$, E. Soldani$^{c}$, S. Sosio$^{f,l}$, S. Spataro$^{f,l}$, E. Tskhadadze$^{c,g}$, S. Verma$^{i}$, R. Wheadon$^{f}$, L. Yan$^{f}$}

\address{$^{a}$Institute of High Energy Physics, Chinese Academy of Sciences, Beijing, China \\
$^{b}$INFN, Sezione di Ferrara, Ferrara, Italy \\
$^{c}$INFN, Laboratori Nazionali di Frascati, Frascati (Roma), Italy \\
$^{d}$INFN, Sezione di Perugia, Perugia, Italy \\
$^{e}$INFN, Sezione di Roma, Roma, Italy \\
$^{f}$INFN, Sezione di Torino, Torino, Italy \\
$^{g}$Joint Institute for Nuclear Research, Dubna, Russia \\
$^{h}$Politecnico di Torino, Dipartimento di Elettronica e Telecomunicazioni, Torino, Italy \\
$^{i}$Universit\`a di Ferrara, Dipartimento di Fisica, Ferrara, Italy \\
$^{k}$Universit\`a di Perugia, Dipartimento di Fisica e Geologia, Perugia, Italy \\
$^{l}$Universit\`a di Torino, Dipartimento di Fisica, Torino, Italy}

\maketitle

\begin{history}
\published{Day Month Year}
\end{history}

\begin{abstract}
  The Cylindrical GEM-Inner Tracker (CGEM-IT) is the upgrade of the internal tracking system of the BESIII experiment. It consists of three layers of cylindrically-shaped triple GEMs, with important innovations with respect to the existing GEM detectors, in order to achieve the best performance with the lowest material budget. It will be the first cylindrical GEM running with analog readout inside a $1$T magnetic field. The simultaneous measurement of both the deposited charge and the signal time will permit to use a combination of two algorithms to evaluate the spatial position of the charged tracks inside the CGEM-IT: the {\it charge centroid} and the {\it micro time projection chamber} modes. They are complementary and can cope with the asymmetry of the electron avalanche when running in magnetic field and with non-orthogonal incident tracks. To evaluate the behaviour under different working settings, both planar chambers and the first cylindrical prototype have been tested during various test beams at CERN with $150$ GeV/$c$ muons and pions. This paper reports the results obtained with the two reconstruction methods and a comparison between the planar and cylindrical chambers.
\end{abstract}

\section{Introduction}
\begin{figure}[htbp]
\centerline{
\includegraphics[width=.25\textwidth]{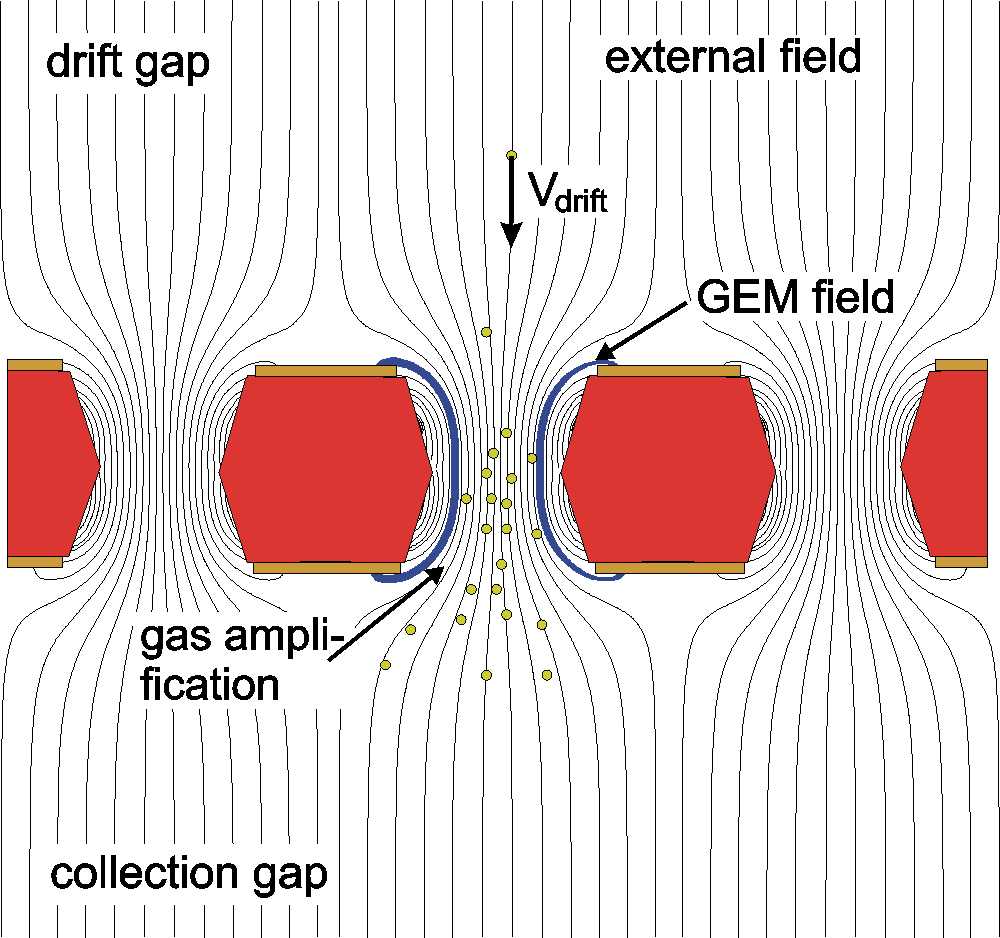}
\qquad
\includegraphics[width=.4\textwidth]{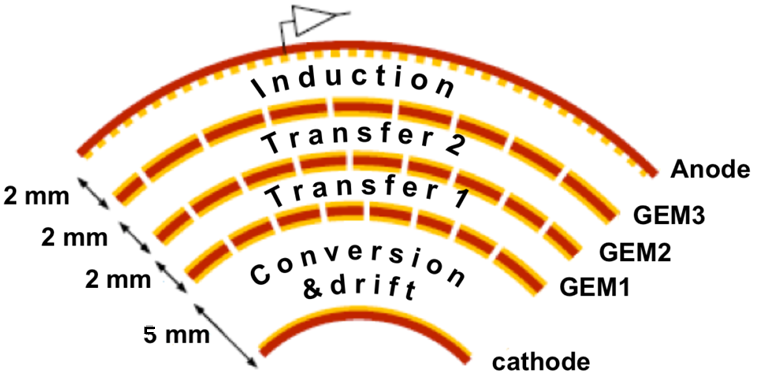}
}
\caption{{\it left}: drift field lines inside the GEM holes; {\it right}: the scheme of the cylindrical triple GEM structure, with electrodes and gas gaps. \label{fig:gem}}
\end{figure}

The CGEM-IT is a tracking detector consisting of cylindrically-shaped triple GEMs, which will serve as the new Inner Tracker of the BESIII experiment\cite{besiii}. \\
\, \\
The Gas Electron Multiplier (GEM) technology was invented by Sauli in 1997\cite{sauli}. It is adopted in the amplification stage of gas trackers to produce high gain values with a low discharge rates\cite{bachmann}. 
In a gas tracker, a charged particle crossing the gas-filled volume produces ionization, creating clusters of electrons and ions. An electric field is applied to drift the electrons towards a region with higher electric field, where they undergo avalanche multiplication. The reached gain is sufficient to allow their detection on the anode. In wire-based trackers, the region of high electric field is created by wires. In the GEM case, it consists in a kapton foil, copper coated on both sides, pierced with thousands of double conical holes of $50/70$ $\mu$m inner/outer diameter. When a voltage of some hundred of Volts is applied between the two copper layers, due to their dimension, an electric field of some tens of kV/cm is created inside the holes (see Fig.~\ref{fig:gem}, {\it left}). As the multiplication wires, the GEM foils are immersed in a drift electric field which guides the electrons inside the holes, where the avalanche multiplication takes place. Gains up to $10^4$ are reachable with moderate voltages. Moreover, with a series of three GEM foils instead of just one between anode and cathode it is possible to obtain the same or higher gain values operating the GEM layers with lower voltages, thus decreasing the probability of discharges\cite{bachmann}.
In the case of the CGEM-IT the triple GEMs are cylindrically molded: a sketch is reported in Fig.~\ref{fig:gem} ({\it right}). The very first cylindrical triple GEM detector belongs to the experiment KLOE-2 (Frascati)\cite{kloe2}. The CGEM-IT of BESIII exploits the legacy of KLOE-2, borrowing for example the construction procedure from it, with important innovations:
\begin{itemlist}
\item[1)] on the mechanical side, in order to minimize the material budget:
  \begin{itemlist}
  \item[i.] Rohacell 31 instead of Honeycomb as support of the anode/cathode;
  \item[ii.] permaglass rings only outside the active area.
  \end{itemlist}
\item[2)] on the electronical side, in order to obtain the best performances:
  \begin{itemlist}
  \item[i.] development of a dedicated ASIC: the TIGER\cite{tiger};
  \item[ii.] jagged strips, to decrease the inter-strip capacitance of around $30\%$\cite{isatipp14}.
  \end{itemlist}
\end{itemlist}
The CGEM-IT is going to be part of the BESIII tracking system. The current tracking detector is the Main Drift Chamber (MDC). It consists of an Inner and an Outer Chamber, which are independent and share the same He-based gas mixture. During the almost 10 years of data taking of BESIII, the inner layers of the MDC were exposed to an always higher radiation dose, as the BEPCII e$^+$e$^-$ collider, which hosts the experiment, increased its luminosity to the design value of $1 \cdot 10^{33}$ cm$^{-2}$ s$^{-1}$, in $2016$. This produced an aging problem and the necessity to find a solution to cope with a possible breakdown of the inner chamber; on the other side, this also gave the opportunity to upgrade the inner tracker with a new technology: the GEM. The most important new feature of the CGEM-IT is the analog readout in an environment with $1$T magnetic field. This peculiarity will make the CGEM-IT a unique detector in its field. The anode of the CGEM-IT is segmented in axial and tilted strips, to give a three dimensional measurement on each layer, and each strip measures both the deposited charge and the time of the signal, providing these two quantities to the reconstruction algorithms. 

\section{Position reconstruction methods}
\begin{figure}[!hbp]
\centerline{\includegraphics[width=.55\textwidth]{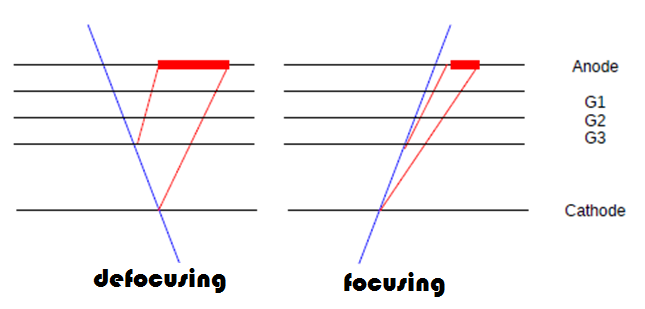}} \label{fig:focdefoc}
\caption{Two scenarios are possible with magnetic field and inclined tracks: {\it focusing}, when the Lorentz and track angle are concordant, and {\it defocusing}, when they are discordant. In the former case the charge distribution is bigger, in the latter it is smaller.}
\end{figure}

Being a tracker, the first goal of the CGEM-IT is a precise position determination, with $130$ $\mu$m resolution in $xy$ plane and $1$ mm on the $z$ coordinate, along the beam direction. In particular, the CGEM-IT is expected to improve the $z$ determination and the secondary vertex position reconstruction w.r.t. the MDC. \\
In order to achieve this goal, two reconstruction algorithms have been implemented: the charge centroid and the micro time projection chanber ($\mu-$TPC). \\
The {\it charge centroid} calculates the $x$ position of the particle on the anode plane with a weighted average of the firing strip $x_i$ positions by the deposited charge $q_i$
\begin{equation} \label{eq:cc}
< x > = \frac{\sum_i x_i q_i}{\sum_i q_i}.
\end{equation}
The {\it $\mu-$TPC} uses the time information recorded on each strip and by knowing the drift velocity it infers the $z_i$ position of the primary ionization: the drift gap works as a {\it micro} ($5$ mm) TPC, the $x_i-z_i$ positions are fitted with a straight line and the $x$ track position in the middle of the gap is calculated
\begin{equation} \label{eq:mutpc}
 x = \frac{\frac{gap}{2} - b}{a}.
\end{equation}
The charge centroid mode is best suited in the case of Gaussian charge distribution and with cluster size $> 1$. The $\mu-$TPC has the best resolution at higher cluster size (more firing strips, which means more points for the straight line fit) and works fine in the case of non-Gaussian shapes. When considering the various track scenarios, the Gaussian shape of the charge distribution is obtained when tracks orthogonal to the GEM plane and with no magnetic field are considered. In this case the only relevant physical effect is the diffusion due to the gas. It enlarges the charge cloud, but does not change its shape. When inclined tracks are considered, the charge distribution shows a tail and the shape is no longer Gaussian. On the other side, if the magnetic field is switched on, considering orthogonal tracks, there is a shift in the position of the charge distribution, due to the Lorentz force, and a modification in the shape of the electron cloud. When the two cases are combined, i.e. we have inclined tracks inside a magnetic field, two scenarios may arise depending on the Lorentz and the inclination angles, as shown in Fig.~\ref{fig:focdefoc}

\section{Test beam results}
\begin{figure}[htbp]
\centerline{
\includegraphics[width=.45\textwidth]{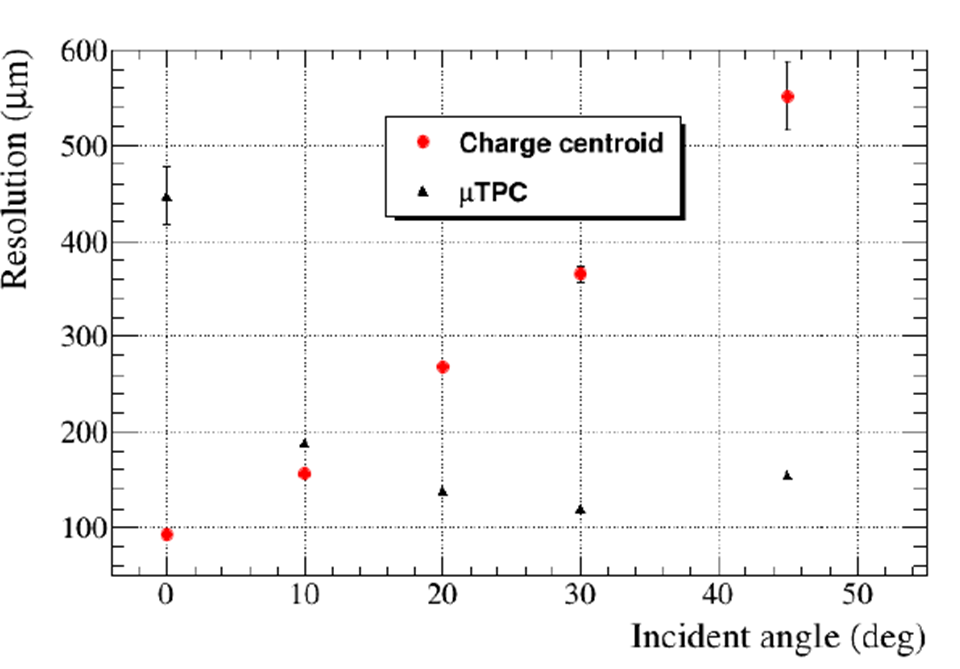}
\qquad
\includegraphics[width=.45\textwidth]{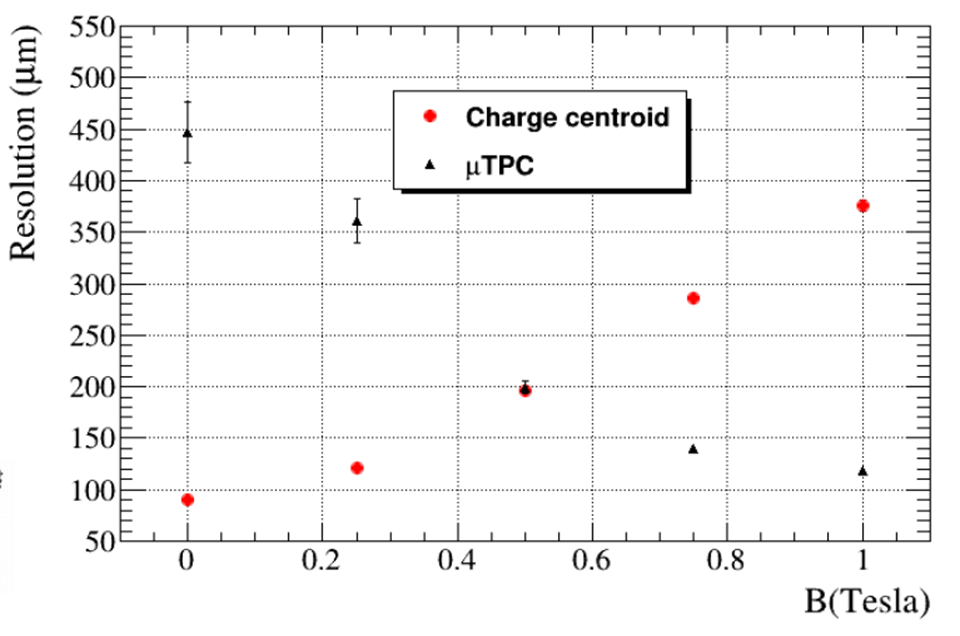}
}
\caption{{\it left}: $B = 0$, resolution {\it vs} incident angle; {\it right}: orthogonal tracks, resolution {\it vs} magnetic field\protect\cite{rictipp17}. \label{fig:resolutions}}
\end{figure}

\begin{figure}[htbp]
\centerline{\includegraphics[width=.55\textwidth]{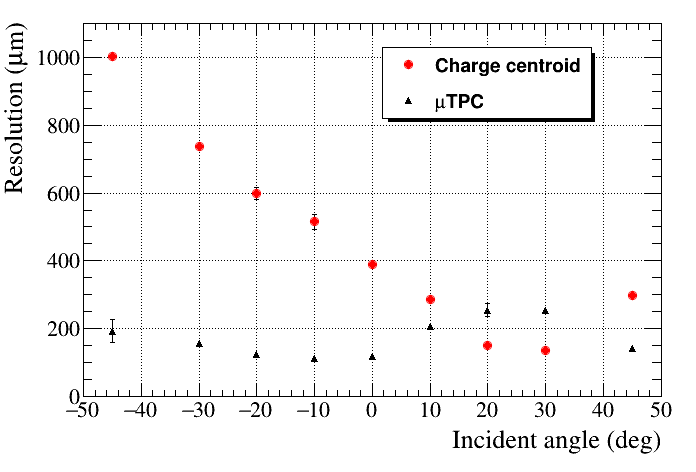}}
\caption{$B \neq 0$, resolution {\it vs} track incident angle\protect\cite{rictipp17}.\label{fig:angleon_Bon}}
\end{figure}

\begin{figure}[!h]
\centerline{\includegraphics[width=.55\textwidth]{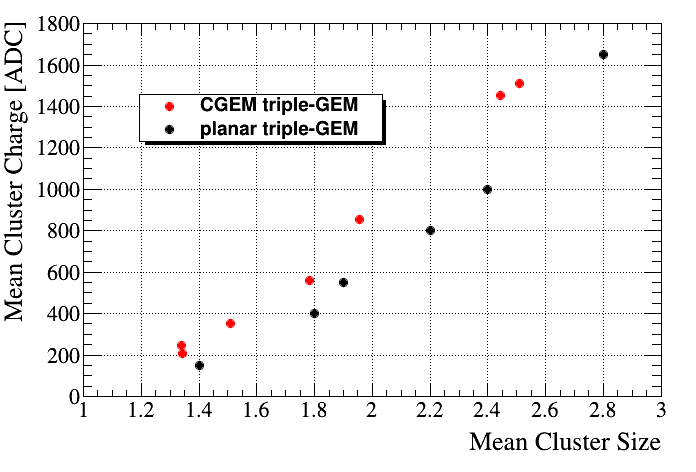}}
\caption{Deposited charge (ADC) {\it vs} cluster size in the CGEM\protect\cite{rictipp17}. \label{fig:cgem}}
\end{figure}

To explore the two algorithms capabilities, as well as to test the chambers in different working conditions, various test beams have been performed with both planar and cylindrical chambers. Here, the results obtained at the H4 line of SPS (CERN) will be shown. The planar chambers were $10 \times 10$ cm$^2$ triple GEM, with two orthogonal views on the anode (strip pitch $650$ $\mu$m), filled with Ar:CO$_2$ ($70:30$) and Ar:iC$_4$H$_{10}$ ($90:10$) gas mixtures. The cylindrical chamber was the first cylindrical prototype, resembling the central layer of the final CGEM-IT, with $x$ and $v$ views, filled with Ar:CO$_2$ ($70:30$) gas mixture. All the chambers were read by APV-25 ASIC chips.\\
The planar chambers were exposed to a muon beam ($150$ GeV/$c$), with and without a rotation, with and without magnetic field. Figures \ref{fig:resolutions} {\it left} and {\it right} show the obtained resolutions {\it vs} the incident angle ({\it left}) and the magnetic field ({\it right}). 
The charge centroid resolution is better than $100$ $\mu$m for orthogonal tracks and without magnetic field, but worsens as the incident angle or the magnetic field strength increase. In these situations, where the number of firing strips is larger, the $\mu-$TPC resolution improves. Figure \ref{fig:angleon_Bon} shows the resolution w.r.t the inclination angle in the case with magnetic field $1$T, which will be experienced in BESIII. 
The focusing and de-focusing regions are clearly visible as in the first one the charge centroid is better than the $\mu-$TPC while in the second one (with a larger number of firing strips) it is the opposite. The combination of the two methods will be able to provide the desired resolution everywhere.
\, \\
The cylindrical chamber was exposed to the pion beam ($150$ GeV/$c$) without magnetic field and for orthogonal tracks. The collected charge as a function of the cluster size (at different gain values) is shown in Fig.~\ref{fig:cgem}. 
The behaviour of the planar chamber is overlaid to the one of the cylindrical GEM: they show the same linearity. This assures the applicability of the reconstruction algorithms developed for the planar chambers also to the cylindrical ones: the obtained resolutions are compatible within tens of microns. A resolution of $110$ $\mu$m was obtained with the charge centroid. \\
\newpage
The complete CGEM-IT is under construction and the commissioning is foreseen in 2018.
\section*{Acknowledgments}
The research leading to these results has been performed within the BESIIICGEM Project, funded by the European Commission in the call H2020-MSCA-RISE-2014.

\end{document}